\documentstyle[emulateapj,psfig]{article}
\def\fz{photometric redshifts }
\def\br{BR1202-07}

\def\cm2{cm$^{-2}$}

\def\lya{Ly$\alpha$ }

\slugcomment{Submitted to AJ}

\lefthead{Fontana et al.}
\righthead{}

\begin{document}

\title{Photometric redshifts and 
selection of high redshift galaxies in the NTT and Hubble Deep Fields.
}

\author{Adriano Fontana}
\affil{Osservatorio Astronomico di Roma, Via dell' Osservatorio 2,
        I--00040, Monteporzio, Italy}
\author{Sandro D'Odorico}
\affil{ESO, Karl-Schwarzschildstr. 2, Garching bei M\"unchen,
         D--85748}
\author{Francesco Poli,Emanuele Giallongo}
\affil{Osservatorio Astronomico di Roma, Via dell' Osservatorio 2,
        I--00040, Monteporzio, Italy}
\author{Stephane Arnouts}
\affil{ESO, Karl-Schwarzschildstr. 2, Garching bei M\"unchen,
         D--85748}
\author{Stefano Cristiani}
\affil{Space Telescope European Coordinating Facility, ESO, Karl-Schwarzschildstr. 2, Garching bei M\"unchen,
         D--85748}
\author{Alan Moorwood}
\affil{ESO, Karl-Schwarzschildstr. 2, Garching bei M\"unchen,
         D--85748}
\author{Paolo Saracco}
\affil{Osservatorio Astronomico di Brera,
22055 Merate (LC) Italy}

\date{} 
\begin{abstract} 

We present and compare in this paper new photometric redshift catalogs
of the galaxies in three public fields: the NTT Deep Field, the HDF--N
and the HDF--S. In the case of the NTT Deep Field, we present here a
new photometric catalog, obtained by combining the existing BVrI and
JK$s$ with new deep U observations acquired with NTT-SUSI2, and which
includes also the contiguous field centered on the $z_{em}=4.7$ quasar
BR1202-07.

Photometric redshifts have been obtained for the whole sample (NTTDF +
HDF--N + HDF--S), by adopting a $\chi^2$ minimization technique on a
spectral library drawn from the Bruzual and Charlot synthesis models,
with the addition of dust and intergalactic absorption.  The accuracy,
determined from  125 galaxies with known spectroscopic
redshifts, is $\sigma _z\sim 0.08 (0.3)$ in the redshift intervals
$z=0-1.5 (1.5-3.5)$.

The global redshift distribution of I--selected galaxies shows a
distinct peak at intermediate redshifts, $z\simeq 0.6$ at $I_{AB}\leq
26$ and $z\simeq0.8$ at $I_{AB}\leq 27.5$ followed by a tail extending
to $z\simeq6$. Systematic differences exist among the fields, most
notably the HDF--S which contains a much smaller number of galaxies at
$z\simeq 0.9$ and at $z\geq4.5$ than the HDF--N.  We also present for
the first time the redshift distribution of the total IR-selected
sample to faint limits ($Ks \leq 21$ and $J\leq22$). It is found that
the number density of galaxies at $1.25 < z < 1.5$ is $\simeq 0.1
$arcmin$^{-2}$ at $J\leq21$ and $\simeq 1. $arcmin$^{-2}$ at
$J\leq22$, and drops to $\simeq 0.3 $arcmin$^{-2}$ (at $J\leq22$) at
$1.5 < z < 2$.

The HDFs data sets are used to compare the different results from
color selection criteria and photometric redshifts in detecting
galaxies in the redshift range $3.5\leq z\leq4.5$.  Photometric
redshifts predict a number of high $z$ candidates in both the HDF--N
and HDF--S that is nearly 2 times larger than color selection
criteria, and it is shown that this is primarily due to the inclusion
of dusty models that were discarded in the original color selection
criteria by Madau et al 1998. In several cases, the selection of these
objects is made possible by the additional constraints from the IR
bands.  This effect partially reflect the poor spectral sampling of
the HDF filter set, and is not present in ground--based observations
where a $R-I\leq 0.5$ color selection criteria may be applied.

Finally, it is shown that galactic M stars may mimic $z>5$ candidates
in the HDF filter set and that the 4 brightest candidates at $z>5$ in
the HDF-S are indeed most likely M stars. 

The data and photometric redshift catalogs presented here are 
available on line at {\sf http://www.mporzio.astro.it/HIGHZ}.

\end{abstract}

\section{Introduction}
Starting with the pioneering studies  of Kron (1980) and Tyson (1988),
one of the main targets of deep imaging surveys has been the
observation and  study of galaxies at high redshift. The
combination of deep imaging and systematic spectroscopic follow--up
has led to the discovery that the bulk of galaxies at the
spectroscopic limit is made of blue star--forming galaxies at $z\leq
1-1.5$, where most of the stars of the present--day Universe appear to
form (see Koo and Kron 1992, Ellis 1998 for extensive reviews).

A new approach recently developed to access the earlier epochs of
galaxy formation relies on deep multicolor surveys, where
multiband images are taken with a complete set of standard broad--band
filters in order to cover the overall spectrum of the galaxy and to
discriminate the populations at different redshifts.  In this case,
the multiband catalogs are used to define sharp color criteria that
select high redshift galaxy candidates (Steidel et al 1995, Madau et
al 1996, Madau et al 1998, M98 hereafter, Fontana et al. 1996,
Giallongo et al 1998, G98 hereafter).  The very successful
results of the spectroscopic follow--up of color selected samples
(Steidel et al 1996, Steidel et al 1999) witness the reliability of the
method.

However, this technique doesn't exploit all the information contained
in multiband catalogs, that allow to sample the spectral shape of each
galaxy over a very extended wavelength range.  With this aim, the ``
photometric redshift'' approach is rapidly becoming common practice.
The concept that the redshift of a galaxy is reflected in its
broad--band colors dates back to the early 60's (Baum 1962), and the
existence of a definite relation between $z$ and magnitude has been
demonstrated by Koo (1985). However, extensive applications of this
method have been developed only recently in order to analyze  new sets of
deep multicolor images. Following the original approach of Koo,
methods have been applied to the HDF-N that make use of the observed
redshift--magnitude relation (Connolly et al 1997, Wang et al 1998) 
derived from the spectroscopic subsample of the HDF--N. A different
class of methods must be used when no large spectroscopic follow--up
has been obtained yet, or when the technique is to be extended beyond
the redshift range covered by the spectroscopic sample. In these
cases, a spectral library is used to compute galaxy colors at any
redshift, and a matching technique is used to obtain the best--fitting
redshift.  With different implementations, this method has been used on
HDF--N (Sawicky 1997, Fernandez--Soto et al 1999,  Benitez et al 1999,
Arnouts et al 1999a) and ground--based data (G98, Fontana et al 1999a,
Pello et al 1999).

Of paramount importance for the development of these applications have
been the public surveys carried out over the last few years, primarily
the HDF--N (Williams et al 1996) and HDF--S (Casertano et al 2000),
since the superb photometric quality of the data and the extensive
follow--up observations enable a large number of scientific issues to
be addressed. Large ground--based telescopes have also been used to
provide multicolor data with an extended wavelength coverage, often to
support deep observations from space facilities. Examples are the VLT
observations of the HDFS--NICMOS field (Fontana et al 1999a) and the
NTT observations of the whole HDF--S (Da Costa et al 1998) and of the
AXAF Deep Field (Rengelink 1998).  In this context, the field of the
$z=4.7$ quasar BR1202-07 was one of the first where deep multicolor
observations have been collected from ground to obtain catalogs
suitable for determining photometric redshifts. First observations in
the BVrI bands were used to search for galaxies at $z\geq4$ in the QSO
surroundings, and led to the identification of a star--forming
companion to the QSO (Fontana et al 1996, see also Hu et al 1996,
Petitjean et al 1996, Fontana et al 1998), providing the first
successful example of the ``drop-out'' technique at $z\geq4$.  From the
same set of data, a photometric redhift distribution was obtained and
discussed in G98.  A much deeper field in the same BVrI bands was
obtained in 1997 with the SUSI1 imager at the NTT, and named ``NTT
Deep Field'' (Arnout et al 1999b). Wider images covering both fields
have been obtained in J and K$s$ during the commissioning phase of the
SOFI infrared spectro--imager at the NTT (Saracco et al. 1999), and also made
publicly available.

In this paper, we present new U band observations of the BR1202
and NTT Deep fields and describe the procedures used to
produce the final UBVrIJK$s$ catalog in both fields (sect. 2).  The
photometric redshift technique is described in sect. 3 and applied to
the NTT catalogs and to the public catalogs of the HDF--N and
HDF--S. The main properties of the resulting photometric redshift
catalogs are described in sect.4. In sect. 5, we focus on the
selection of high redshift galaxies to discuss the differences of this
approach with respect to the Lyman Lymit selection and the possible
contamination by foreground M stars.

The photometric redshift catalogs presented here have been used to
address several scientific issues in the high redshift Universe,
ranging from the evolution of the luminosity density and the number of
massive galaxies already assembled at early epochs (Fontana et al
1999b) to the evolution of galaxy sizes (Poli et al 1999, Giallongo
et al 2000).

The data and photometric redshift catalogs are also available on line at the
WEB site {\sf http://www.mporzio.astro.it/HIGHZ}.

\section {The data}

\subsection{Observations and data reduction on the BR1202-07 fields}

The new data used in this paper are deep UBVrIJK$s$ images taken with
the ESO--NTT telescope on the field of the high redshift quasar
BR1202-07 ($z_{em}=4.7$).

The B,V, Gunn $r$ and I observations were taken with the NTT SUSI
imager during two separate observing runs, on 23--26 Apr 1995 and
during the spring of 1997 respectively. The pointings were chosen in
order to obtain two slightly overlapping fields, with the first field
(hereafter named BR1202) centered on the quasar and the second
( NTT Deep Field, NTTDF hereafter) 100 arcsec south of it, resulting in
a total field of 2.3'$\times$4.'  Since observations were obtained
with two slightly different instrumental setups, the two fields have
been analyzed independently and the objects in the overlapping region
have been used as a final test of the photometric catalog.  The BVrI
observations and data reduction of the BR1202 field are discussed in
Fontana et al 1996 and G98.  The NTTDF was observed in service
mode by the NTT team in February through April 1997, and the data
obtained are publicly available at {\sf
http://www.hq.eso.org/science/ndf}.  Details of the observations,
the data reduction and the photometric catalog are discussed in
Arnouts et al 1999b. Image quality in both fields is quite good, with
sub--arcsec seeing in all the final images.  An image of the final field
in the $r$ band is shown in Fig~\ref{image}.

The J and K$s$ observations were obtained with the SOFI
spectro--imager at the ESO--NTT (Moorwood et al 1998a) in March 1998.
The field of view of this instrument is considerably larger
(5'$\times$5') than the SUSI one so both fields could be
observed simultaneously.  The data reduction and calibration of the
images are described in Saracco et al 1999 and publicly available
at the ESO website. 

We have extracted from the full format J and K$s$ frames two
sub--images corresponding to the two SUSI fields described above, by
locating a number of corresponding bright sources and applying a
linear resampling to the images. Since this operation added a further
rebin to the original pixels, altering the noise statistics, we
estimated the final limiting magnitude of the IR images by computing
the $\sigma$--clipped standard deviation of the intensity of the
sky--background as measured in random positions of the rebinned image
with a 1$''$ aperture (we have verified that the same procedure on the
original image provides the same standard deviation). The resulting
$3\sigma$ limiting Johnson magnitudes are $J\leq 23.4$ and $Ks \leq
21.7$ respectively.

These data have been complemented with deep U images obtained with the
new SUSI2 imager at ESO--NTT. This instrument is equipped with a mosaic
of two $4k\times 2k$ CCD chips with a field of view of $5.5' \times 5.5'$
that matches the SOFI one. Observations were done with the same
pointing as the SOFI observations, with the two former SUSI fields
both falling on the same chip.  Data were obtained during two clear
nights on 21 and 24 April 1998. A total of 25200 second of integration time
were obtained, split into individual exposures of 1200-1500 s
each. Observations were done in dark sky under moderate seeing
conditions (about 1.4$''$ on average).  The instrument performance was
somewhat reduced with respect to the nominal one  because of residual
parasitic light entering the detector, that produced an additional
background pattern with an intensity of about 40\% of the sky
background in the U band. Several exposures of this scattered light
were taken during a bad weather night (25 April 1998). The pattern
proved to be constant and could therefore be subtracted from the individual raw
images before further processing. Aside from this aspect, the data
reduction followed the usual steps for deep images acquired in
dither mode. It has been performed with the automated DITHER
pipeline developed for the SUSI2 images at the Rome Observatory. It
makes use of standard Midas commands and of SExtractor (Bertin and
Arnouts 1998) to flat-field, align and coadd the images to produce a
final image together with its absolute variance map (i.e. the variance
in ADU of each pixel in the frame).

A set of spectrophotometric standard stars have been observed at
several airmasses during both nights to calibrate the U images in the AB
system.  We estimate that the final accuracy of this procedure is of
the order of 0.1 {\it mags}. Again, two smaller images rebinned and
aligned to the BV$r$I frames were extracted, and their zeropoints were
computed in order to match the U magnitude of the brightest objects.

For the reader's convenience, we summarize in Table1 the main
characteristics of these observations and compare them with those
of the HDF--N and HDF--S, which cover similar areas and 
will also be used in the following sections.

\subsection{The photometric catalogues  in the BR1202 fields}

Object detection and flux measurement have been performed with the
SExtractor software.  In the NTTDF we have repeated the procedure used
in Arnouts et al 1999b by using the sum of the BV$r$I frames, that have
comparable seeing. The numbering, position and BVrI magnitudes of the
NTTDF provided here are therefore identical to those published in
Arnouts et al 1999b.  In the case of the BR1202 field we have used the
$r$ image, that has the best seeing, to detect and deblend the
objects, partly for consistency with G98 and partly because the
addition of the BVI images - that have either worse seeing or less
depth - did not result in an improved object detection.  Since the aim
of this work is to derive photometric redshifts of objects that are
significantly above the detection threshold, the differences in the
procedures adopted for object detection are not important, although
they must be kept in mind when using the data. In particular, only the
NTTDF has been used in Fontana et al 1999b to measure the evolution of
the rest--frame UV luminosity density, using the I--selected sample.
We have also performed an independent object detection on the K$s$
images alone to ensure that all the galaxies at $Ks<21$ were included
in the catalogs.  The only object detected in the IR images but not in
the optical is the methane brown dwarfs already discussed in a
separate paper (Cuby et al 1999).

We have first obtained a reference magnitude $m_r$ in the
$r$ band for each object in both fields by obtaining both isophotal
and aperture magnitudes in a 2x FWHM aperture.  The isophotal magnitude
was used for the brighter objects, i.e. for those objects where the
isophotal radius is larger than the aperture one. For fainter objects,
an aperture correction to $5''$ has been estimated on bright stars and
applied to correct the 2x FWHM aperture magnitude. Although this
procedure is strictly valid for star-like objects only, it has been
shown to be a good approximation for faint galaxies (Smail et
al. 1995).

To obtain total magnitudes in the other bands we have first measured
aperture magnitudes in a fixed circular area corresponding to 2x
FWHM of the detection frame. Total magnitudes in each band $i$ have 
then been computed as 
$m_i = m_i(2FWHM)- m_r(2FWHM)+ m_r $, i.e. the observed aperture colors with
respect to the reference $r$ band
have been normalized
to the total  $r$  magnitude  $m_r$ as obtained above.
This procedure in principle ensures that colors - that are the 
quantities used in estimating the redshifts - are measured in the same 
physical region at any wavelength. In ground--based images
this procedure is complicated by the different seeing of the
images in the various bands. To compensate for this, we have degraded
the $r$ reference frame to the seeing of each band and computed 
the aperture colors $m_i(2FWHM)- m_r(2FWHM)$ on the images
with the same seeing. To give an idea of the importance
of this effect, we have  found roughly that each 0.1$''$ of difference
in seeing translates to about a 0.1 {\it mags} difference in
the aperture color, although the exact value depends obviously on
the object morphology.

The zeropoints were corrected for galactic absorption assuming $E(B-V)
= 0.02$ (Burstein \& Heiles 1982) with $\delta U=0.095$, $\delta
B=0.084$, $\delta V=0.063$, $\delta R=0.05$, $\delta I=0.038$.  We
summarize in Table 1 the zeropoints of the photometric systems adopted
for the catalogs, namely AB for all the HDF catalogs and for
the $U$ band of the NTT fields, and Johnson for all the other bands of
the NTT fields. Since we have not applied any color term to the
Johnson magnitudes of the catalogs,, the resulting magnitudes are
calibrated in the ``natural'' system defined by our instrumental
passbands (see Fontana et al 1996).

We have used the overlapping section of the two fields to compare the
two photometric catalogs - that have been taken with slightly
different instrumental setups and at different epochs.  The
differences found ($\Delta B = + 0.05$, $\Delta V = -.135$, $\Delta r
= -0.05$, $\Delta I = -.12$ ) are close to  the limit of the quoted
uncertainties on the zeropoints. Since the BR1202 field has been
observed under more controlled photometric conditions, we have applied this
set of small corrections to the NTTDF catalog to make the two catalogs
fully consistent. However, the differences may be indicative of the
actual uncertainties associated with the photometry on deep exposures
from ground--based telescopes, that typically result from the stacking
of several exposures obtained on different nights.
We will discuss in Sect3.2 the possible effects 
of these uncertainties on the \fz obtained.

The final catalogs produced (along with the photometric redshifts, see
sect. 3) are reported in Table 2 and are available on line 
at {\sf http://www.mporzio.astro.it/HIGHZ}. This table contains also
the half--light radius measured with the morphological analysis
described in Poli et al 1999.

\subsection{The final galaxy samples}

To produce the final catalogs on which photometric redshifts are
computed, we have identified and removed from the final galaxy sample
obvious bright stars using the
CLASS\_STAR parameter provided by SExtractor. A threshold CLASS\_STAR$
< 0.9$ has been set (at $\leq24$ in the NTTDF and $\leq23.5$ in the
BR1202 field), on the basis of the comparison between ground--based
and HST data (Arnouts et al 1999b).  We have also made use of the more
detailed morphological analysis by Poli et al (1999) to identify stars
down to $I=24.5$ on the NTTDF only.

We have applied our photometric redshift code also to the HDF-N and
HDF--S with the IR observations obtained at Kitt Peak (Dickinson 1998)
and at NTT--SOFI (Da Costa et al, 1998), respectively.  For the HDF--N
we have used the multicolor catalog published by Fernandez--Soto et al
(1999), which uses an optimal technique to match the optical and IR
images that have a quite different seeing. A similar catalog for the
HDF--S has been provided by the same authors and is available at the
Stony Brook WEB site {\sf
http://www.ess.sunysb.edu/astro/hfds/home.html}.  The Stony Brook
catalog of the HDF--S contains also the NTT optical magnitudes
(UBVRI) for each object, obtained from the public data of the EIS deep
survey.  Very little difference on the photometric redshift is found
when the NTT data are used or not, which is due to the lower
statistical weight (compared to WFPC data) that is associated with
them. For this reason, we have used only WFPC bands in the optical and
NTT--SOFI in the IR, for consistency with the HDF--N.  In both cases
we have clipped the outer regions of the frame with lower S/N.
Obvious stars have been excluded at $I_{AB}\leq 25.5$ in the HDF-S
again on the basis of the SExtractor CLASS\_STAR$ < 0.9$ morphological
classification.

Also the complete set of HDF photometric redshift catalogs is
available at {\sf http://www.mporzio.astro.it/HIGHZ}

\section {Photometric redshifts}

\subsection {The technique}
 
Much emphasis has been  recently placed  on the choice of the spectral 
library used to compute galaxy colors as a function of redshift. 
Several authors (e.g. Fernandez--Soto et al 1999, Benitez 1999)
use observed spectral templates (e.g. Coleman et al 1980).  
This has the obvious advantage 
of only relying on a compact set of empirically determined data, 
but these must be extended in the UV and IR regions 
using synthetic models and augmented with bluer templates 
to represent the vast fraction of blue galaxies that dominate 
the counts at faint magnitudes.  
An even more empirical approach is to determine a set of autofunctions 
from the spectroscopic control sample (Csabai et al 2000), from
which a set of template spectra can be derived.

Our recipe for photometric redshifts is based on the use of spectral
synthesis models, at present the GISSEL library by Bruzual \& Charlot.
At the cost of an additional computational effort, these models allow
- at least in principle - to take into account the spectral evolution
of galaxies at high redshift.  We will show below that these models
are able to provide a redshift accuracy which is at least 
equivalent to more
empirical choices.  However, the main reason for following this
approach is the gain in physical information on the properties of high
redshift galaxies that can be obtained by comparing their observed
spectral energy distributions with the model predictions.  From the
input parameters used to draw the best fitting solution for each
galaxy we can indeed obtain estimates of the main physical quantities
of each galaxy in the sample: age of the last major starburst, mass
already assembled in stars and dust content.  Even taking into account
the known degeneracies among some of the input parameters, one may use
this information in a statistical way to further constrain galaxy
evolution.  Applications of this technique have already been presented
in G98 and we refer to forthcoming papers for further applications of
this method.

In practice, we have used a Miller-Scalo initial mass function (IMF)
including stars in the $0.1 < M < 65$ M$_{\odot}$ range (the effect of
different choices of the IMF has been shown in G98).  We have
considered the cases of metallicity $Z= Z_{\odot}$, $Z=0.2 Z_{\odot}$
and as low as $Z=0.02 Z_{\odot}$ with a weak correlation between age
and metallicity.  For these models, we span over a grid of $e-folding$
star formation time-scales $\tau$ (from $0.3$ Gyrs to constant) and
ages to produce the evolution of different spectral types (the
complete grid adopted is listed in G98). Since we are aware that the
smoothly declining star--formation histories that we have used are a
simplified rendition of the actual histories experienced by real
galaxies, we have added a few models with additional bursts of star
formation.  In practice, we have overimposed short-lived burst (0.3
Gyr of timescale) to two evolving models (0.3 and 5 Gyr of timescale),
starting at an age of 3 Gyr. In each model, we let a substantial
fraction of the final mass in the galaxy to be assembled during the
burst: 10\%, 30\% or 50\% in the former and 10\%, 20\% or 30\% in the
latter (for higher fractions the light from the starburst would
dominate that from the underlying galaxy at any wavelength). {\it A
posteriori}, we have verified that ``multiple burst'' models account
for about 16\% of the best--fitting models. In 30\% of them, removing
the multiple bursts from the library changes the best fit by more than
$\Delta z=0.05$, and in 14 \% of the cases by more than $\Delta
z=0.2$, but always less than $\Delta z=0.5$. In all these cases the
burst models are selected to match the contemporary presence of a flat
star--forming UV continuum and a large 4000\AA break with a
significant slope of the IR continuum, that is difficult to match with
the simple exponential laws (see also the discussion on the Scalo $b$
parameter in G98).  Although the spectroscopic sample is too small to
provide a meaningful statistics, we note that a marginal improvement
of the photometric redshift accuracy is usually obtained with the
``multiple burst'' models.  More important is the effect that these
models have on the parameters estimated for the best fitting
spectrum. For instance, the stellar mass in each galaxy is found lower
by typically 20\% (but up to 60\%) at fixed redhift, because of the
contribution of post--AGB stars (that have a low M/L) to the near--IR
flux.

We have also added to all spectral synthesis model the reddening
produced by internal dust (SMC, Pei 1992 and Calzetti, 1997) and Lyman
series absorption produced by the intergalactic medium (Madau
1995). As a further improvement with respect to G98, we have added
[OII] and H$\alpha$ emission lines by inverting the Kennicutt 1994
relations for the relevant IMF.  The \lya emission is less
straightforward to implement, since the spectral properties at the
\lya frequency observed in spectroscopic samples range from strong
emission lines to damped absorption systems.  As a tradeoff, we have
removed from the unattenuated spectrum the stellar absorption features
at the \lya frequency.

At any redshift, galaxies are allowed to have any age smaller than the
Hubble time at that redshift ($\Omega =1$ and $H_0 = 50$ km
s$^{-1}$Mpc$^{-1}$ have been adopted throughout the paper).  The
choice of the cosmological parameters are not important, since only
colors are effectively used in deriving the photometric redshifts.

From the large dataset ($\simeq 5\times 10^5$) of ``simulated galaxies'' 
thus  produced we compute the expected fluxes $F_{template,i}$ in any
filter $i$ by convolving the theoretical spectrum with the 
normalized transmission curve of the telescope+detector+filter system
(see Pickles 1998 for the formulae adopted).

Finally,  a $\chi ^2$--minimization procedure has been
applied to find the best--fitting spectral template to the observed
fluxes.  For each template $t$ at any redshift $z$ we first minimize
\begin{equation}
\chi ^2_{t,z} = \sum_i \left[ {F_{observed,i}-s\cdot F_{template,i}
\over \sigma _i} \right]^2
\end{equation}
with respect to the scaling factor $s$ where $F_{observed,i}$ is the
flux observed in a given filter $i$, $\sigma _i$ is its uncertainty
and the sum is over the filters used, and then we identify the
best--fitting solution with the lowest $\chi ^2_{t,z}$.  The scaling
factor $s$ is applied to the input parameters of the input spectrum to
compute all the rest--frame quantities, such as absolute magnitudes or
stellar masses. The Stony Brook catalogs of the HDF--N and HDF--S
provide  $F_{observed,i}$ and $\sigma_i$ already in physical units
(Jy), and are therefore directly usable in Eq. 1. In the two NTT
fields, where catalogs are given in standard magnitudes
$m_i$ (zeropointed to the Johnson system), $F_{observed,i}$ is given by
$F_{observed,i}= 10^{-0.4\times(m_i-48.59-C_i)}$, where $C_i$ is the
conversion factor from the AB to the Johnson systems, found by
convolving an A0V stellar spectrum of the Pickels 1998 library with the
telescope+detector+filter efficiency.

As is well known, the conversion between fluxes and magnitude is
not well behaved as the noise level is approached, which raises the
question of how non-detections are handled in Eq. 1. In the cases of
the two HDFs, the Stony Brook catalogs provides very small or even
negative fluxes for objects that are non detected (like ``U--dropout''
or faint galaxies with no IR detection), so that Eq. 1 can be directly
applied.  In the NTT fields, we have followed the common practice of
defining a magnitude limit at a given $\sigma$ level, and computed
$\chi ^2_{t,z}$ according to the following recipe: {\it a)} we have
discarded from the sum in Eq.1 the bandpasses where {\it both} the
observed galaxy flux and the models fall below the limiting magnitude;
{\it b)} we have included in the sum the bandpasses where the models
lay above the limit: to weight properly the non--detection, the object
has assigned zero flux and noise equal to the $1 \sigma$ magnitude
limit.

The former approach  is closer to the physical
definition of ``measure'' and statistically cleaner, but relies on an
accurate estimate of the noise level at very low fluxes, which in real
images may be dominated by non-poissonian effects (such as residuals
of flat--fielding, pixel correlation and accuracy of sky--subtraction)
that may be difficult to quantify. The latter may be useful either on
public data with conventional formats or when conservative upper
limits on fluxes are more appropriate to the data available.  It is
therefore interesting to investigate the effects of the different
criteria on the final photometric redshifts. To this purpose, we have
converted the Stony Brook HDF--N catalog into magnitudes, and defined
the $1\sigma$ limits in the various bands by locating the magnitude at
which $\Delta m \simeq 1.08$. We compare the photometric redshifts
obtained in the original "flux" format and in the conventional
"magnitude" one in Fig~\ref{uplim}, differentiating galaxies in the
three samples used in Fontana et al 1999b: $K\le 21, I_{AB}\le 26,
I_{AB}\le 27.5$ It is seen that very modest differences arise only in
the faintest sample, with a few galaxies migrating from $z\simeq 0.5$ to
$z=3-4$ and even less in the opposite way (which makes the choice of
the ``flux'' format more conservative), and with some scatter in the
$z=1-2$ region, where the different treatment of the IR band plays a
role.

\subsection {Comparison with spectroscopic samples}
This procedure has been tested on a sample of 125 galaxies with
spectroscopic redshifts and multicolor photometry. Of these, 112 come
from the Cohen et al (2000) compilation of the HDF--N (we have
included only objects that fall in the inner region of the frame, as
for the whole optical catalog, see Sect. 2.3), 3 are taken from
the Stony Brook public catalogs of the HDF--S, while the remaining 10
are taken from the VLT spectroscopic follow--up of ``dropout''
galaxies in the HDF--S and AXAF deep fields (Cristiani et al 2000).
The results are shown in fig.~\ref{zspe}, where we differentiate
between objects with secure or uncertain redshift, following the
classification by Cohen et al (2000). 
 
If we consider only the objects with secure redshift, and remove the
object at $z\simeq2.9$ with $z_{phot}\simeq0.2$, for a total 101
objects, the accuracy is $\sigma_z \simeq 0.08$ at $0 <z < 1.5$ and
$\sigma_z \simeq 0.3$ at $z>2$, that we keep as reference value.  If
we take all the galaxies in the sample, and remove only the three
objects with clearly discrepant redshift,  the accuracy raise to
$\sigma_z \simeq 0.1$ at $0 <z < 1.5$ and $\sigma_z \simeq 0.32$ at
$z>2$.  If we compute a 3-sigma clipped rms we obtain an accuracy
$\Delta z = 0.09$ over the whole redshift range, with 111 objects out
of 125 (i.e. 89\%), and $\Delta z = 0.07$ in the $0 <z < 1.5$ interval
(85 objects out of 91, i.e. 93\%). In any case, the relative
precision $\sigma [(z_{phot}-z_{spec})/(1+z_{spec})]$ is always of the
order of 0.05 in the $z<1.5$ sample, as in most of similar
analysis (Cohen et al 2000).
 
A 5\% systematic underestimate $< \Delta z > \simeq -0.14$ appears to
exist at $z>2$ in the spectroscopic sample, similar to that found by a
recent revised version of the photometric redshifts based on the
Coleman spectra ($< \Delta z > \simeq 0.15$, Lanzetta 2000), but with
opposite sign.
 
We will explore in future works how this underestimate can be reduced
with a different choice of the GISSEL grid. 
In particular, we are collecting a much wider spectroscopic sample in
different fields (D'Odorico et al, in preparation, and within the
``K20 project'', P.I. Cimatti) and with a variety of bandwidths to
perform a more accurate calibration of the spectral library,
especially in the  range $2<z<4$, where most of the observed
data come from the rest--frame UV spectra.

Undoubtedly, the availability of only one major spectroscopic control
sample (the HDF-N) is a major limitation that prevents to some extent
a full evaluation of all the systematics involved in \fz.

It is worth reminding that at least 4 different ingredients are needed
to pin down an accurate recipe for photometric redshifts: i) the
photometric catalog (which depends on the accuracy of data processing
and calibration and on the procedure used to measure colors); ii) the
knowledge of the response curve of the telescope+filters+detector
system; iii) the spectral library used and iv) the matching algorithm
used.  Despite this, most of the current debate is focused on the best
choice of the spectral library and the relative influence of the
different components in a typical survey is far from being assessed,
especially in the case of ground--based observations.

Some hints on the expected biases may be obtained by Monte Carlo
simulations. However, these typically assume errors to be Gaussian
distributed which may be an oversimplification since, at the very
bright and very faint limits, the uncertainties involved in background
estimation and flat--fielding may dominate the errors. In
Fernandez--Soto et al 1999 and in Arnouts et al 1999a Monte Carlo simulations have
been used to show that photometric redshifts in the HDF--N catalogs
are stable with respect to noise fluctuation down to about
$I_{F814W}\simeq28.5$.   Arnouts et al 1999a showed that the maximum number of
misidentifications (i.e. shift in $z$ exceeding $0.5$) is below 30\%
in the $I_{F814W}\simeq28.5$ catalogs.  We adopt here a brighter limit
$I_{F814W}\leq27.5$ ( much brighter than the $I_{F814W}$ limiting
magnitude), that is ultimately set by the depth of the bluest images
($U_{F300W}$), so that a minimum $(U-I)_{AB}\geq1$ is still
measurable. At this magnitude the misidentification are obviously much
lower (see fig 4 of Arnouts et al 1999a). By analogy, we set an upper limit to $I_{AB}
\le 26$ in the NTTDF.

As a first test of the effect of systematic differences, we have
compared the \fz obtained in the NTTDF with the two different
photometric calibrations (see Sect. 2.2). Fig.~\ref{ZP} (upper panel)
shows the histogram of the scatter between the \fz $z^{'}_{phot}$
obtained with the ``original'' calibration and the \fz $z_{phot}$
resulting with the adopted calibration.  We first note that, for most
of the objects, the difference found is within the expected accuracy of
\fz~. Objects with clearly discrepant redshifts are typically faint 
and amount to only $\simeq 3\%$ of the total, a result that is
consistent with the Monte Carlo simulation of Arnouts et al 1999a,
when scaled do the depth of the NTTDF.  A close scrutiny of the
$z^{'}_{phot} - z_{phot}$ relation (fig.~\ref{ZP}, lower panel) shows
that the systematic changes to the zeropoints induce systematic
offsets in the \fz that vary with redshift as the spectral breaks
pass across the various filters. These changes marginally affect
the selection of high redshift galaxies in the NTTDF: two
objects at $z\simeq 2.8$ in the final catalog are assigned at $z'\simeq0.5$
with the original calibration (out of 40 in the $2.5<z<3.5$ bin),
compensated by two objects that are moved from $z'\simeq2.8$
to much lower $z$. One galaxy (out of 14) is added to the $3.5<z<4.5$ bin
with the adopted calibration. The major statistical effect is due to 
5 objects that move from $z'\simeq2.3$ to $z\simeq2.6$, slightly
changing the predicted values in the UV luminosity density in the
bins adopted in Fontana et al 1999b.

In the long run, we expect that extended spectroscopic follow-up of
multicolor deep fields with different filter sets and in different
regions of the sky will allow a detailed evaluation of the different
systematics involved.

\section {Redhift distributions}
We discuss in this section the fundamental properties of the redshift
distributions in the catalogs. Beyond being the basis for several
scientific studies  - as discussed in the introduction - these
distributions can be used to compare the predictions of photometric
redshift analyses in different fields and with different instrumental
setups.  Given the limited size of the fields (the HDFs and
the NTTDF have areas of 4-4.8 arcmin$^2$, respectively) small scale
fluctuations are expected to lead to discrepancies in the redshift
distributions.  At the same time, photometric redshifts may still be
regarded as an experimental technique and small uncertainities in the
photometric calibration or inadequacy of the spectral templates in the
different filter sets may also lead to small but systematic effects in
the resulting redshifts. At the present stage, it is not posssible to
fully disentangle these effects, pending full spectroscopic
follow-ups and independent checks of the image calibrations.  However,
the comparison among the results in different fields may be regarded
as an estimate of the total uncertainities in this kind of
analysis. For an evaluation of its effects on some scientific
topics see Fontana et al 1999b.

Fig~\ref{nz_br1202} shows the redshift distribution in the catalogs of
the two fields around \br, separately (we remind the reader that these
catalogs have different depths). The distribution of
Fig~\ref{nz_br1202}{\it b} of the BR1202 field is similar to the one
published in Giallongo et al 98, where no U and J band and poorer K data
were available. We find here the
typical characteristics of redshift distributions at faint limits i.e most of
the galaxies are located at $0.5 < z < 1$ with a tail extending to greater
$z$.

Although the comparison between the $r$--selected distribution of the
BR1202 field and the I--selected distribution of the NTTDF is not
straightforward, we note that the tail of high redshift galaxies is -
as expected - more pronounced in the deeper sample. We also note that
the position of the peak in the redshift distribution that is skewed
to lower redshifts in the NTTDF compared to the BR1202 field, opposite
to what is naively expected in a deeper sample. This is, however,
consistent with the observed number densities in the two fields - the
NTTDF being richer in relatively bright galaxies - and emphasizes the
need for much wider surveys.

The average redshift distributions at faint limits can be obtained by
merging the NTTDF and the two HDFs. Fig~\ref{nz_I}. shows the
I-limited distributions, at $I_{AB} \le 26$ (for a total of 973
objects), resulting from the combination of NTTDF and the two HDFs,
and at $I_{AB} \le 27.5$ (1361 objects), from the two HDFs alone.

The global
redshift distribution  shows again the well known
features of a distinct peak at intermediate redshifts,
 $z\simeq 0.6$ for the brighter and  $z\simeq 0.8$ for the fainter
followed by a tail extending to the highest redshifts compatible with the 
I selection criteria.

A close scrutiny of the individual distributions reveals systematic
differences amongst the fields. At both magnitude limits, the HDF--S
has a much smaller number of galaxies at $z\simeq 0.9$ than the HDF--N.
Since the filter set used in both observations is identical, this effect is
not likely to be an artefact of the  photometric redshifts. 

Finally, we use the combined data sets of NTTDF and HDFs to build the
redshift distrbution in IR selected samples shown in
Fig~\ref{nz_IR}.  The basic scientific output of these distributions
has been discussed in Fontana et al 1999b, where it is shown that the
redshift distribution of K--band selected galaxies provides a direct
estimate of the number of massive galaxies already assembled at high
redshift. We provide here a more detailed view of the redshift
distribution in J and K--selected samples at different magnitude
limits. These distributions may be useful for tuning extended
spectroscopic surveys with the new IR spectrographs that are being
implemented at 8--m telescopes.  These surveys are designed to improve
the knowledge of the large scale structures and emission
line features of galaxies in the range $1 < z < 2$, where
spectroscopic measurements with optical spectrographs are difficult
because of the lack of suitable features.  We detail the
observed number densities at different J and Ks limits in Table~\ref{ircounts}.
We find that the number
density of galaxies at $1. < z < 1.5$ is $\simeq
1 $arcmin$^{-2}$ at $J\leq21$ and $\simeq 3.8 $arcmin$^{-2}$ at
$J\leq22$, and drops to $\simeq 0.3 $arcmin$^{-2}$ (at $J\leq22$) at
$1.5 < z < 2$.

\section{The selection of high redshift galaxies}

\subsection {Galaxies at $3.5<z<4.5$}
 
The search for extremely high redshift galaxies is one of the most
interesting applications of deep multicolor imaging. The so--called
``dropout'' technique has gained widespread popularity due to its
simplicity and high success rate in detecting galaxies at
$z\geq2.8$. However, it is worth remembering that this method relies
on the contemporary presence of {\it two} distinct features: the sharp
cutoff provided by the intrinsic Lyman Limit (at $z\leq 3.5$) or by
the IGM Lyman$\alpha$ absorption (at higher $z$) {\it and} the flat
star--forming continuum at longer wavelengths, that samples the
rest--frame extreme UV spectral region. The latter criterium is
necessary to discriminate against interlopers (e.g. intermediate $z$
elliptical or dusty galaxies, or even late--type stars) that may also show a
dramatic drop in their rest--frame near UV.  In a nutshell, high
redshift galaxies appear as objects that are ``blue in the red and red in
the blue''. An important consequence of this approach is that the
``dropout'' technique is essentially a highly conservative technique
that is designed to minimize the fraction of interlopers, but that is
biased against even moderately dust--reddened galaxies: as shown in
Pettini et al 1997, even a relatively weak dust absorption
($E(B-V)\simeq0.15$ with an SMC extinction law) would prevent the
detection of galaxies at $z>3$ in their well-studied $UGR$ color
plane.

Given the strength of the IGM color feature, color selection criteria
and photometric redshifts provide comparable results when the filter
set ensures an appropriate coverage of both features. This is
illustrated at $z\simeq3$ in the VLT observations of the HDFS NICMOS
field (Fontana et al 1999a) and at higher $z$ in the previous version
of the BR1202 catalog (G98), where a $r-I\leq0.2$ requirement was
explicitly chosen to detect galaxies at $4\leq z\leq4.5$. A similar
requirement $R-I\leq0.6 $ has been adopted by Steidel et al 1999 to
improve the rejection of low $z$ interlopers in their first systematic
spectroscopic follow up of $z\geq4$ candidates.

Unfortunately, a similar agreement between photometric redshifts and
color selection criteria does not hold in the case of galaxies at $3.5
\leq z \leq 4.5$ in the HDF--N and HDF--S. As a result, the estimate of the UV
luminosity density obtained with color selection criteria
($\phi_{1500} \simeq 5.0 \times
10^{25}$~erg~s$^{-1}$~Hz$^{-1}$~Mpc$^{-3}$, M98) is significantly
different from that obtained with photometric redshifts (($\phi_{1500}
\simeq 10^{26}$~erg~s$^{-1}$~Hz$^{-1}$~Mpc$^{-3}$, Pascarelle et al
1998, Fontana et al 1999b). This is primarily due to the much larger
number of candidates selected by the latter technique (roughly 2 times
larger).  This difference appears rather puzzling at first glance,
because both methods have been applied to the same data set (albeit on
different catalogs) and the M98 selection criterium has been defined
using a synthesis spectral library similar to the one used
here.

To investigate the origin of this discrepancy, we reproduce in
fig~\ref{madau} the $B_{450}-V_{606}$ vs $V_{606}-I_{814}$ plane
adopted by M98, whose selection criteria requires
galaxy candidates to fall within the polygonal region shown in
fig~\ref{madau} (solid line), {\it and} to have $V_{606}\leq27.7$. The
latter requirement ensures a proper detection of the ``$B$--dropout''
and is therefore ultimately set by the depth of the $B_{450}$ frame.
We also plot in fig~\ref{madau} the observed galaxy colors of the
joined HDF--N + HDF--S catalogs, differentiating galaxies at different
photometric redshifts. At variance with the 
M98 catalog, our data are limited at $I_{814}\leq27.5$, that corresponds
to a rest--frame wavelength of $\simeq 1600$~\AA.

The following conclusions result from this comparison:
 
- Most of the galaxies lying within the M98 polygonal region have a
photometric redshift falling in the $3.5 \leq z \leq 4.5$ range. The
few exception are in any case at $z\geq 3.3$, in agreement with the
expected scatter of photometric redshifts.  This confirms our
statement that the strengths of the ``dropout'' signature is strong
enough to provide a unique redshift assignment.  Conversely, about
50\% (i.e. 29 out of 54) of the galaxies with $3.5 \leq z_{phot} \leq
4.5$ fall within the M98 area (we include also the two objects at the
left of the M98 area).

- Of the remaining fraction (25 out of 54) of the $3.5 \leq z_{phot}
\leq 4.5$ candidates, most are distributed in a ``belt'' around the
M98 area, that is indeed occupied by high $z$ theoretical models (see
Fig. 5 of M98) but was excluded to minimize the contamination by lower
redshift interlopers. 

- finally, other objects lay progressively apart from
the M98 region, filling the bridge toward the color position of
galaxies at $z>4.5$, with progressively larger $V_{606}-I_{814}$ and
smaller $B_{450}-V_{606}$ with respect to what is required by the M98
approach. These "$z=4$ outliers" are typically faint in $V_{606}$ and
have a very large $V_{606}-I_{814}\geq1$.
 
The brightest among these "$z=4$--outliers" is HDF--N--3259--652, that
is marked by a large circle in fig~\ref{madau} and whose spectral
distribution and best fitting spectrum is shown in
fig~\ref{hdf173}. This object has a large $B_{450}-V_{606}$ because it
is undetected in $B_{450}$ and is the brightest object among this
class (these objects are undetected in $B_{450}$, so that the
$B_{450}-V_{606}$ should be read as lower limits), and has a
spectroscopic redshift of 4.58 and a photometric redshift $z =
4.26$. It is fitted by the spectrum of a star--forming galaxy with a
significant amount of dust ($E(B-V)\simeq0.15$ with an SMC extinction
law) to account for the large $B_{450}-V_{606}$ and $V_{606}-I_{814}$
observed colors.  As can be seen from fig~\ref{hdf173}, the unique
identification of this galaxy is made possible  by the additional
constraints provided by the IR bands: the best-fitting solution when
the J, H and K$s$ band are not used corresponds indeed to an evolved
galaxy at $z_{phot} =0.57$, consistent with the M98 plot (as for all
the objects in the spectroscopic sample, the photometric redshift does
not depend on how non-detection in the B band is treated).

To evaluate the effect of the IR bands on the photometric redshift
selection of high $z$ galaxies, and in particular on these ``z-4
outliers'', we have obtained a fit on the HDF--N and HDF--S samples
after removal of the IR bands. The comparison between the 7--bands and
the 4--bands photometric redshifts is shown in fig~\ref{noIR}. It is
shown that, while all the objects selected with the M98 criteria are
assigned at $3.5<z<4.5$ also with only 4 optical bands, a substantial
number of the "$z=4$ outliers" is placed at $z\leq 1.5$ when the IR
bands are removed. These objects have redshifts $z>4$, and are
typically fitted with significant amounts of dust
[$<E(B-V)>\simeq0.1$]. In all these cases the role of IR bands is to
constrain the fit to relatively small values of $I_{814}-K$, even in
the cases where only upper limits exist in the IR, with the result of
discarding models of evolved or dusty galaxies at lower $z$ that could
easily mimic the red $B_{450}-V_{606}$ and $V_{606}-I_{814}$ optical
colors.

Also the "$z=4$ outliers" that are placed at $z>4$ irrespective of the
inclusion of IR bands (a total of 6 out of 54) are selected among
models with $<E(B-V)>\simeq0.1$, that were present also in the M98
synthetic grid but not included in the selection criteria.  Respect to
the M98 approach, the faint objects with large $V_{606}-I_{814}>1$ and
$B_{450}-V_{606}\leq2$ were also discarded by not only because of
their colors but also because they fade beyond the $V_{606}\leq 27.7$
limit applied by M98.

Not suprisingly, the situation is different in our ground--based
images. If we remove the IR bands from the NTTDF sample, all the
$z>3.5$ candidate are found as well, and all the $z\simeq 4$
candidates by G98 are found by photometric redshifts, and viceversa.
The reason lays in the different spectral sampling of the two filter
sets.  The spectral region covered by the three filters $B_{450}$,
$V_{606}$ and $I_{814}$, where Lyman$\alpha$ blanketing starts to
affect the HDF filter set already at $z\simeq 3.5$, is sampled by 4
different filters ($BVrI$) in the ground based observations used here
or by other groups.  

{\it Summarizing, the $3.5<z< 4.5$ candidates that are selected with
photometric redshifts in the HDFN+S data set and that lay far from the
M98 region are typically predicted to be objects at $z>4$ with
substantial amount of dust, i.e. a class of objects which
multicolor criteria are biased against}. The IR bands are often
required with the HDF filter set to include these objects in the
sample.  A correction of similar amplitude to the M98 criteria
for dusty objects  has been
derived by Steidel et al 1999 using the observed color distribution of
galaxies in their spectroscopic survey at $z\simeq 3$.

{\it It is also worth emphasizing that the differences between the
methods have not been tested with spectroscopic follow--up in a systematic
way}.  It is certainly reassuring that HDF--N--3259--652, that lays
outside the M98 area albeit it is at $z=4.58$, is recovered with the
photometric redshift analysis, but the major differences arise from
the fainter sample, where the constraints from the IR bands are less
stringent and the noise on the observed magnitudes is larger.

Another issue that should not be underestimated concerns the
differences among different algorithms used for source
detection and measure, even on the same data. Based on the same set of
HDF--N and HDF--S images, Casertano et al 2000 find a larger number of
$z\simeq4$ candidates than  are found in the Stony Brook catalogs,
at the same magnitudes and adopting the same M98 criteria, and find
that the HDF--N shows less candidates respect to the HDF--S, contrary
again to what it is found in the Stony Brook catalogs.

We conclude with a comment on fig~\ref{noIR}: we note that, as
expected, the lack of IR bands causes a relatively large scatter in
the $1<z<2$ region, the region where the major spectral breaks fall in
the IR bands. However, the number of completely discrepant redshifts
($\Delta z >1$) is limited. To some extent, this reflects the fact
that for a typical object the S/N ratio of the WFPC bands is always
much higher than that of the IR bands, so that they dominate the
$\chi^2$ even when the IR bands are included.

\subsection {Galaxies at $z>4.5$}
 
It is well known that Galactic stars are one of the main sources of
foreground interlopers that may be identified as high redshift
galaxies.  As discussed in sect. 2.3, we have applied a pure
morphological criterium to select bright stars in the HDF--S and in
the NTTDF, where spectroscopic follow--up is still lacking.  It is
interesting to note that 8 (out of 32) of these ``morphological''
stars are assigned at $z>4$ by our photometric redshift code (see also
Lanzetta 2000).  Most of these objects are very bright ($I_{AB}\leq
23.5$) so that the morphological classification is expected to be very
robust, since high redshift galaxies are compact but resolved in high
quality images, as shown by Giavalisco et al 1996 and Giallongo et al
2000.  However, we have complemented the morphological selection with
a comparison with the stellar library by Pickles 1998, that we have
used as input to our photometric redshift code. We have therefore
compared the $\chi^2_{star}$ obtained for each object in the HDF--S
with the use of the Pickles 1998 library with the $\chi^2_{GISSEL}$
obtained by the ordinary procedure.
 
We find that, while all bright galaxies are very poorly fitted by any
stellar template, with $\chi^2_{star} >> \chi^2_{GISSEL}$ (typically
by a factor of 100), 85\% of the stars selected by morphological
criteria have $\chi^2_{star} \simeq \chi^2_{GISSEL}$ and, in
particular, five of the $z>4$ candidates have $\chi^2_{star} <
\chi^2_{GISSEL}$.  Two typical examples are shown in
fig~\ref{mstar}. It is shown that the high redshift identification
comes from the flat continuum at long wavelengths, coupled with the
large spectral break in the bluer bands - all features typical of high
redshift galaxies (see previous section).  However, they are better
fitted by an M star spectral template because of the relatively gentle
fading - when compared to the abrupt drop due to the IGM at $z>4.5$ -
at blue wavelengths and the negative slope at $\lambda > 10000\AA$
that are both typical of M stars.  We conclude that also the color
properties of these objects support the morphological classification
of faint halo stars.

It is important to emphasize that, once bright M--stars (i.e. objects
that correspond to the morphological criteria discussed above) are
removed, {\it no other convincing candidate at $z>5$ is found in the
HDF--S sample}. In addition to the photometric redshift predictions,
this can be tested directly by looking at the multicolor data: no
object in the Stony Brook catalog of the HDF--S is detected in
$I_{F814W}$ (at $I_{F814W}\leq 27.5$) but not in $V_{F606W}$, contrary
to what is observed in the objects detected so far at $z\geq5$ in the
HDF--N (Lanzetta et al 1996, Spinrad et al 1998). As a result, the UV
luminosity density $\phi_{1400}$ at $z\geq 4.5$ is still poorly
determined, since it changes by a factor of $\simeq 6$ between the
HDF--N and the HDF--S at $z\geq 4.5$ (Fontana et al 1999b).

\section {Summary}

This paper presents the photometric redshift catalogs that have been
used in previous papers to investigate different aspects of the
evolution of galaxies in the high redshift Universe, such as the
history of the UV luminosity density and the number of massive
galaxies already assembled at early epochs (Fontana et al 1999b) and the
evolution of galaxy sizes (Poli et al 1999, Giallongo et al 2000).
 
A new deep multicolor (UBVrIJK$s$) photometric catalog has been produced
of the galaxies in the NTT Deep Field and in the slightly overlapping
field centered on the $z_{em}=4.7$ quasar BR1202-07. This has been
obtained by combining the existing BVrI and JK$s$ images with new, deep,
U band observations of both fields acquired with NTT-SUSI2.
 
We have further presented in this paper the photometric redshift catalog
drawn from this galaxy sample, using a $\chi^2$ minimization technique
based on the Bruzual and Charlot spectral library, with the addition
of dust and intergalactic absorption. We have also presented the
results of applying the same photometric redshift techniqe to public
catalogs of the HDF--N and HDF--S, where a simlar optical-infrared
coverage is available.

The method has been tested on a set of 125 galaxies with known
spectroscopic redshifts in the HDF--N, HDF--S and AXAF fields, with a
resulting accuracy $\sigma _z\sim 0.08 (0.3)$ in the redshift
intervals $z=0-1.5 (1.5-3.5)$.

The global redshift distribution of I--selected galaxies shows a
distinct peak at intermediate redshifts, $z\simeq 0.6$ at $I_{AB}\leq
26$ and $z\simeq0.8$ at $I_{AB}\leq 27.5$ followed by a tail extending
to $z\simeq6$. Systematic differences exist amongst the fields, most notably
the HDF--S contains a much smaller number of galaxies at $z\simeq 0.9$
and at $z\geq4.5$ than the HDF--N.  We have also presented the redshift
distribution of the total IR-selected sample, which may be useful to
tailor the planned surveys with IR spectrographs at large telescopes
that will target the redshift range $1.3 \leq z \leq 2$.  We find that
the number density of galaxies in the redshift range is $\simeq
0.15$arcmin$^{-2}$ at $J\leq21$ and $\simeq 1.3 $arcmin$^{-2}$ at
$J\leq22$.

We have also discussed the different results from applying color
selection criteria and photometric redshifts for detecting galaxies in
the redshift range $3.5\leq z\leq4.5$ using the HDFs data sets.  We
find that photometric redshifts predict a two times larger number of
high $z$ candidates in both the HDF--N and HDF--S and show that this
is primarily due to the inclusion of slightly dusty ($E(B-V)\simeq
0.1$ with SMC extinction law) models that were discarded in the
original color selection criteria conservatively applied by Madau et
al 1998. In several cases, the selection of these objects is made
possible by the additional constraints from the IR bands.  This effect
partially reflect the poor spectral sampling of the HDF filter set,
and is not present in ground--based observations where a $R-I\leq 0.5$
color selection criteria may be applied.

Finally, we show that galactic M stars may mimic $z>5$ candidates
in the HDF filter set and that the 4 brightest candidates at $z>5$ in
the HDF-S are, indeed, most likely to be M stars.  The estimates of the UV
luminosity density $\phi_{1400}$ at $z\geq 4.5$ from these data, when
the selection against halo stars is applied, show that $\phi_{1400}$
changes by a factor of $\simeq 6$ between the HDF--N and the HDF--S
(Fontana et al 1999b).

{\bf Acknowledgments}\\

The paper is based on observations made with: the ESO New Technology
Telescope at the La Silla Observatory (some under the EIS
programs 59.A-9005(A), 60.A-9005(A)), the NASA/ESA Hubble Space
Telescope and the Kitt Peak National Observatory.  The ultraviolet
observations of the NTTDF were performed in SUSI-2 guaranteed time of
the Observatory of Rome in the framework of the ESO-Rome Observatory
agreement for this instrument.

\clearpage
\begin{table*}	
\tablenum{1} 
\caption{Summary of the Observational Data\tablenotemark{1}}  
\label{summary} 
\begin{tabular}{lccccccccc} 
\tableline 
\tableline 
Field name & Size& ~ & Magnitude limit\tablenotemark{2}\\ 
\tableline 

~& (arcmin$^2$)  & U & B& V& Gunn r& I& J& H& K\\ 
\tableline 
 
BR1202 & 4 & 26.9 & 26.5 &26 &25.6 &25  & 23.4 & -- & 21.7\\ 
NTTDF & 4.8 & 26.9 & 27.5 &26.85 &26.5 &26.4  & 23.4 & -- & 21.7\\ 
HDF-N & 4.2 & 28.4\tablenotemark{3} & 29.2\tablenotemark{3} &29.5\tablenotemark{3} &~ &29\tablenotemark{3} & 23.8 & 22.9 & 22.4\\ 
HDF-S & 3.9 & 28.2\tablenotemark{3} & 29\tablenotemark{3} &29.3\tablenotemark{3} &~ &28.7\tablenotemark{3} & 24 & 22.1 & 22\\ 
\tableline 
\end{tabular} 

\tablenotetext{1}{:Note on the photometric system adopted: $U$, F300W, F450W, F606W
and F814W are in AB system: all other data are in the Johnson (i.e. Vega
zeropointed) system.  $J_{AB}$, $H_{AB}$, and $K_{AB}$ magnitudes in
the HDF catalogs have been converted to Johnson applying $J = J_{AB}
-0.87$, $H = H_{AB} -1.34$ and $K = K_{AB} -1.84$. }

\tablenotetext{2}{: at
3$\sigma$. Limiting magnitudes have been estimated from the
photometric catalogs used in the paper, and defined as the typical
value at which $\Delta m = 1.08/3$. Only exceptions are the F606W
and F814W bands, that have been taken with the same criteria
from the STScI public catalogs.}

\tablenotetext{3}{: The F300W,
F450W, F606W and F814W filters of WFPC2 were used for the optical
observations, and magnitudes are given in the AB system}

\end{table*}

\clearpage

\begin{table*}
\tablenum{2} 
\caption{NTTDF and BR1202 photometric catalogs\tablenotemark{1}}  
\begin{tabular}{rcccccccllll}
\tableline
\tableline
 ID\tablenotemark{2}&$\alpha$ [h]&$\delta$ [deg]&U$_{AB}$&B&V&R&I& J& K&$z_{phot}$&$r_{hl}\tablenotemark{3}$\\
 n0001&12:05:18.01&-7:44:40.98&24.74&24.28&23.33&22.85&22.63&21.92&21.01&0.34&0.25\\
 n0002&12:05:18.43&-7:44:40.43&24.72&24.46&23.83&23.06&22.12&20.93&19.43&0.80&0.63\\
 n0003&12:05:18.21&-7:44:39.53&23.35&23.20&22.42&21.79&21.61&21.14&20.28&0.26&0.66\\
 n0005&12:05:17.81&-7:43:33.34&25.19&24.86&24.27&23.96&23.54&22.28&21.76&0.10&~\\
 n0006&12:05:17.79&-7:43:26.36&25.80&25.88&24.43&24.12&23.97&22.49&21.63&0.29&~\\
\tableline
\end{tabular}
\tablenotetext{1}{The complete version of this table is in the electronic
edition of the Journal. The printed edition contains only a sample.}
\tablenotetext{2}{Objects named {\it n0001, n0002} etc, refer to the NTTDF; 
objects named {\it q0001, q0002} refer to objects in the QSO field (BR1202).}
\tablenotetext{3}{Half--light radius, as measured in Poli et al 1999.}
\end{table*}

\clearpage

\begin{table*}
\tablenum{3} 
\label{ircounts}
\caption{Total number and surface densities of galaxies at $z\geq1$ in
the NTTDF + HDF-- + HDF--S sample as a function of redshift. Numbers
within brackets refer to the individual fields, namely NTTDF (upper),
HDF--N (middle) and HDF--S (lower).}
\begin{tabular}{c|cc|cc|cc}
\tableline
\tableline
$z$ & gal   & gal/arcmin$^2$ & gal  & gal/arcmin$^2$ & gal  &  gal/arcmin$^2$ \\ 
\tableline
\tableline
~ & \multicolumn{2}{c|}{J $\le$ 20} &\multicolumn{2}{c|}{J $\le$ 21} &\multicolumn{2}{c}{J $\le$ 22} \\
\tableline
{\it all} 
& 43~~{\scriptsize $\left( \begin{array}{c}16\\16\\11\end{array}\right)$} & 3.08 
& 96~~{\scriptsize $\left( \begin{array}{c}32\\36\\28\end{array}\right)$} & 6.88 
& 196~~{\scriptsize $\left( \begin{array}{c}68\\74\\54\end{array}\right)$} & 14.06 \\

$1<z\le1.25$ 
& 1~~{\scriptsize $\left( \begin{array}{c}0\\1\\0\end{array}\right)$} & 0.07 
& 12~~{\scriptsize $\left( \begin{array}{c}1\\7\\4\end{array}\right)$} & 0.86 
& 38~~{\scriptsize $\left( \begin{array}{c}11\\16\\11\end{array}\right)$} & 2.73 \\

$1.25<z\le1.5$ 
& 0~~{\scriptsize $\left( \begin{array}{c}0\\0\\0\end{array}\right)$} & 0. 
& 1~~{\scriptsize $\left( \begin{array}{c}0\\0\\1\end{array}\right)$} & 0.07 
& 13~~{\scriptsize $\left( \begin{array}{c}2\\4\\7\end{array}\right)$} & 0.93 \\

$1.5<z\le2$ 
& 1~~{\scriptsize $\left( \begin{array}{c}1\\0\\0\end{array}\right)$} & 0.07 
& 1~~{\scriptsize $\left( \begin{array}{c}1\\0\\0\end{array}\right)$} & 0.07 
& 5~~{\scriptsize $\left( \begin{array}{c}2\\2\\1\end{array}\right)$} & 0.36 \\ 

$z>2$ 
& 0~~{\scriptsize $\left( \begin{array}{c}0\\0\\0\end{array}\right)$} & 0 
& 0~~{\scriptsize $\left( \begin{array}{c}0\\0\\0\end{array}\right)$} & 0 
& 2~~{\scriptsize $\left( \begin{array}{c}1\\1\\0\end{array}\right)$} & 0.14 \\

\tableline
\tableline
~ & \multicolumn{2}{c|}{Ks $\le$ 19} &\multicolumn{2}{c|}{Ks $\le$ 20} &\multicolumn{2}{c}{Ks $\le$ 21} \\
\tableline
{\it all} 
& 69~~{\scriptsize $\left( \begin{array}{c}24\\28\\17\end{array}\right)$}  & 4.95 
& 125~~{\scriptsize $\left( \begin{array}{c}43\\44\\38\end{array}\right)$}  & 8.98 
& 247~~{\scriptsize $\left( \begin{array}{c}85\\88\\74\end{array}\right)$}  & 17.72 \\

$1<z\le1.25$ 
&6~~{\scriptsize $\left( \begin{array}{c}0\\4\\2\end{array}\right)$}  & 0.43 
& 24~~{\scriptsize $\left( \begin{array}{c}5\\11\\8\end{array}\right)$}  & 1.72 
& 48~~{\scriptsize $\left( \begin{array}{c}15\\20\\13\end{array}\right)$}  & 3.44 \\

$1.25<z\le1.5$ 
& 1~~{\scriptsize $\left( \begin{array}{c}0\\0\\1\end{array}\right)$}  & 0.07 
& 5~~{\scriptsize $\left( \begin{array}{c}1\\0\\4\end{array}\right)$}  & 0.36 
& 15 ~~{\scriptsize $\left( \begin{array}{c}2\\4\\9\end{array}\right)$}  & 1.08 \\

$1.5<z\le2$ 
& 1~~{\scriptsize $\left( \begin{array}{c}1\\0\\0\end{array}\right)$}  & 0.07 
& 5~~{\scriptsize $\left( \begin{array}{c}1\\3\\1\end{array}\right)$}  & 0.36 
& 17 ~~{\scriptsize $\left( \begin{array}{c}3\\6\\8\end{array}\right)$}  & 1.21 \\

$z>2$ 
& 0~~{\scriptsize $\left( \begin{array}{c}0\\0\\0\end{array}\right)$}  & 0 
& 0~~{\scriptsize $\left( \begin{array}{c}0\\0\\0\end{array}\right)$}  & 0 
& 13 ~~{\scriptsize $\left( \begin{array}{c}8\\2\\3\end{array}\right)$}   & 0.93 \\
\tableline
\tableline
\end{tabular}

\end{table*}

\clearpage
\begin{figure}
\caption{
THIS FIGURE IS AVAILABLE AT http://www.mporzio.astro.it/HIGHZ.
The total field of view around BR1202-07, made by the 
superposition of the two NTT fields discussed here.
Upper field is the BR1202 field. A square marks the position
of the $z=4.7$ QSO. North is up, East left.
}
\label{image}
\end{figure}  

\clearpage
\begin{figure}
\plotone{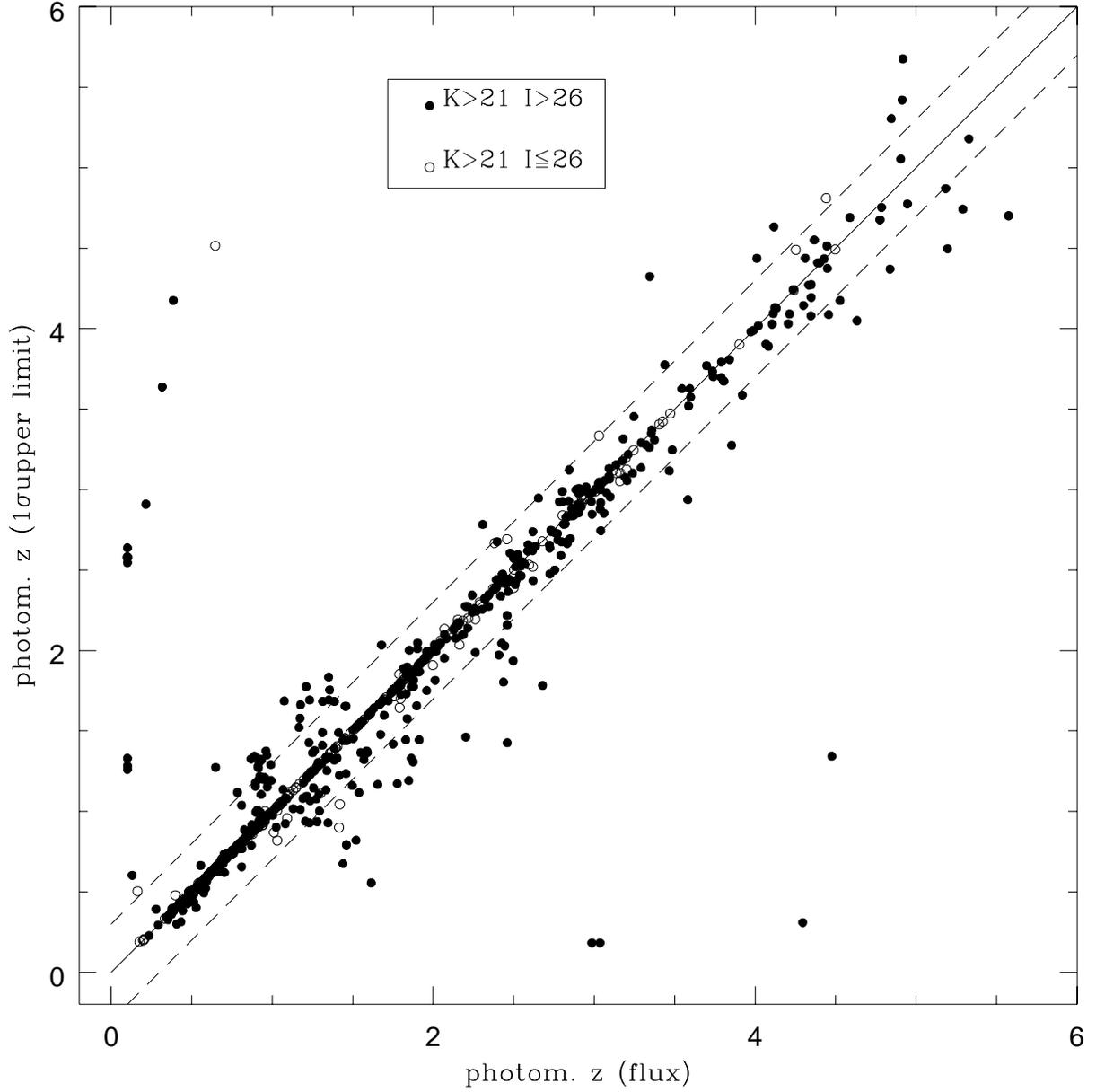}
\caption{ Comparison between the photometric redshifts in the HDF--N 
obtained with the ``flux'' format (on x-axis) used in the paper
and with the ``magnitude'' format with upper limits at $1 \sigma$ level
(see text for full details). The symbols for
objects with different $I_{814}$ magnitudes are given in the legend. Objects
with $K\leq 21$ are shown as crosses, but lay along the bisector line
and are not visible.
  }
\label{uplim}
\end{figure}

\clearpage
\begin{figure}
\plotone{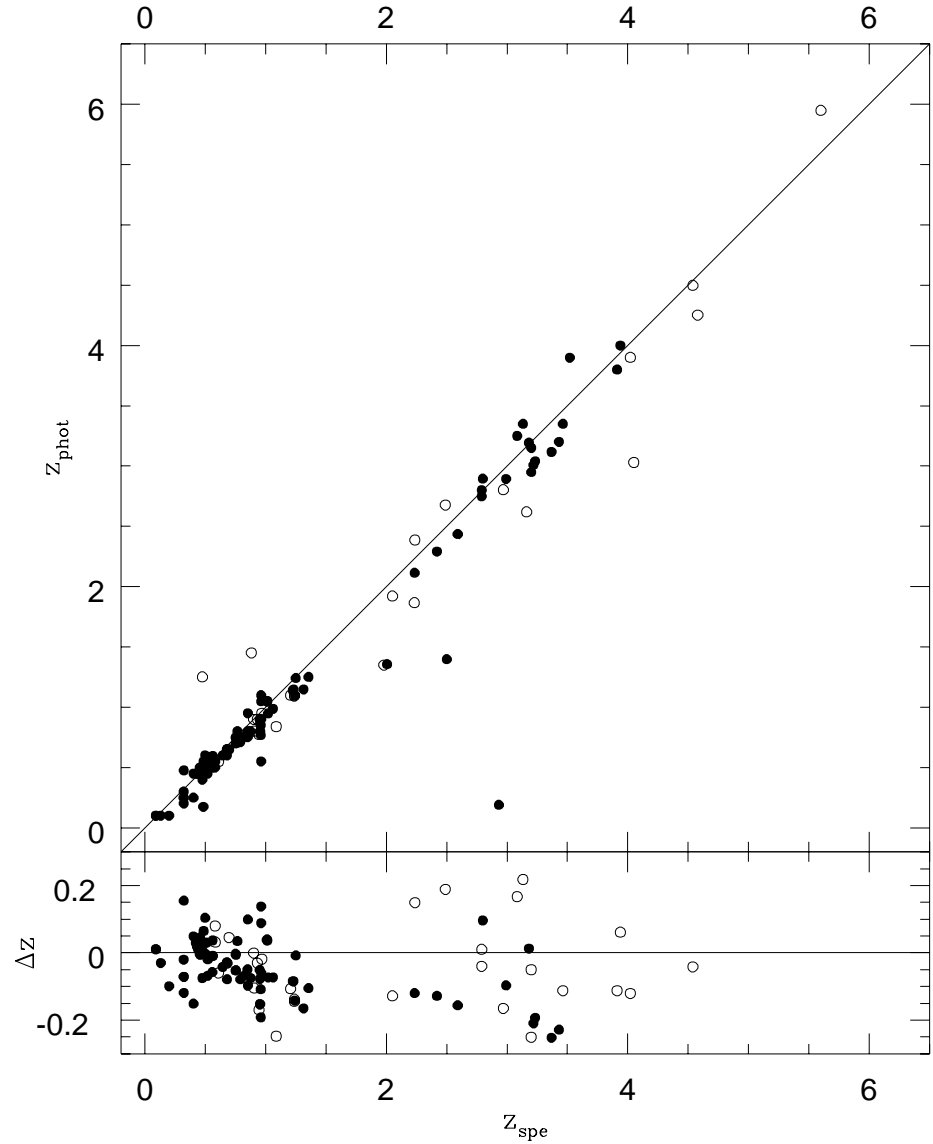}
\caption{
Comparison between the spectroscopic and photometric redshifts
in a sample of 125 galaxies in the HDF--N, HDF--S and AXAF fields.
Filled circles show objects from HDF--S, AXAF and 
with secure redshifts in the HDF--N
(i.e. those  with redshift quality class $=$ 1,2,4 or 6 in Cohen et al 2000),
while empty circles are objects with more uncertain redshift in the HDF-N.
}
\label{zspe}
\end{figure}

\clearpage
\begin{figure}
\plotone{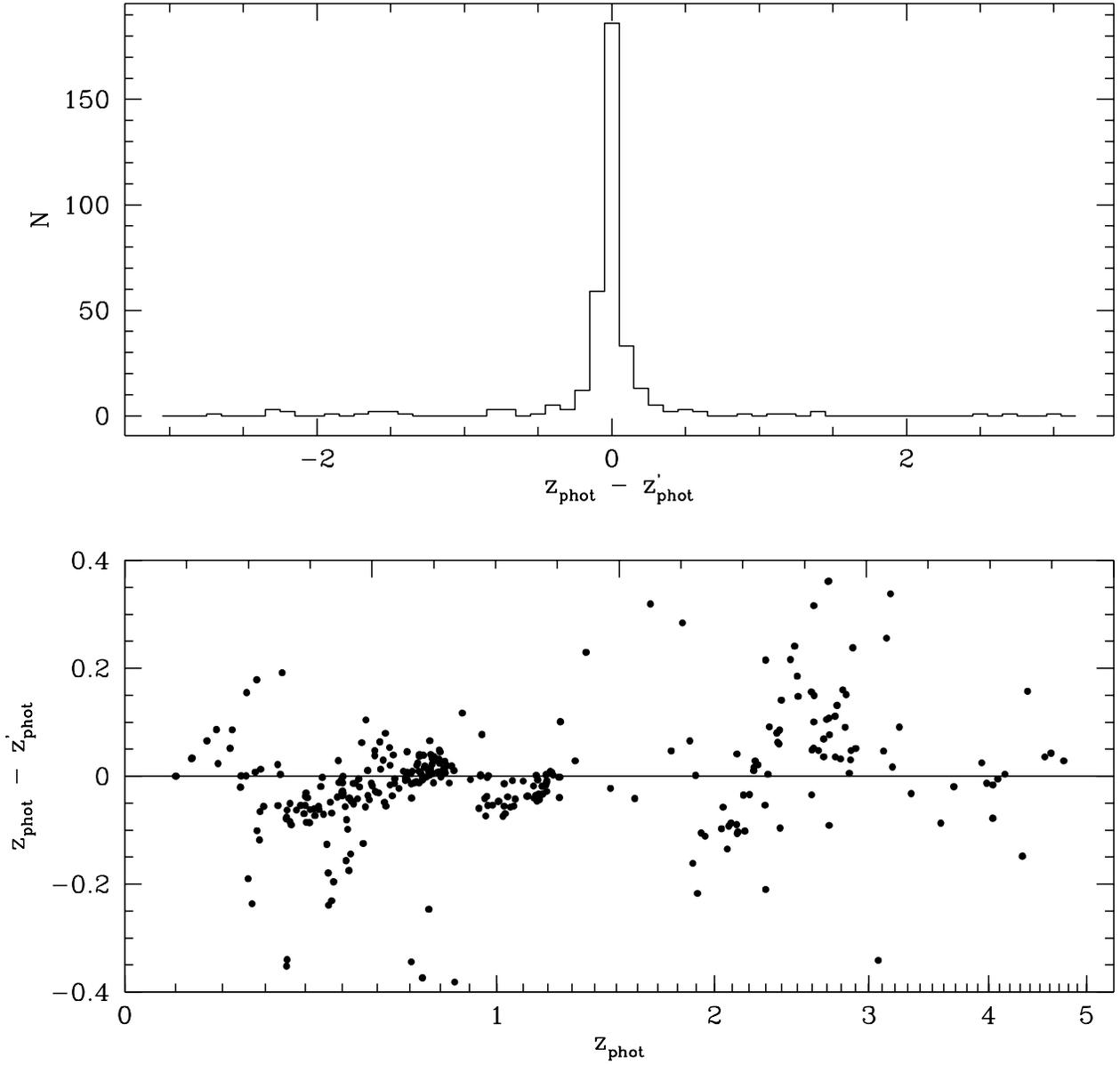}
\caption{ Differences in the \fz estimated on the NTTDF when the two
photometric calibrations are used. $z^{'}_{phot}$ mark the \fz when
the original calibration is used, while $z_{phot}$ are those resulting
from the adopted one.  Upper panel shows the histogram of
$z_{phot}-z^{'}_{phot}$, while the lower one shows
$z_{phot}-z^{'}_{phot}$ as a function of $z_{phot}$.  }
\label{ZP}
\end{figure}

\clearpage
\begin{figure}
\plotone{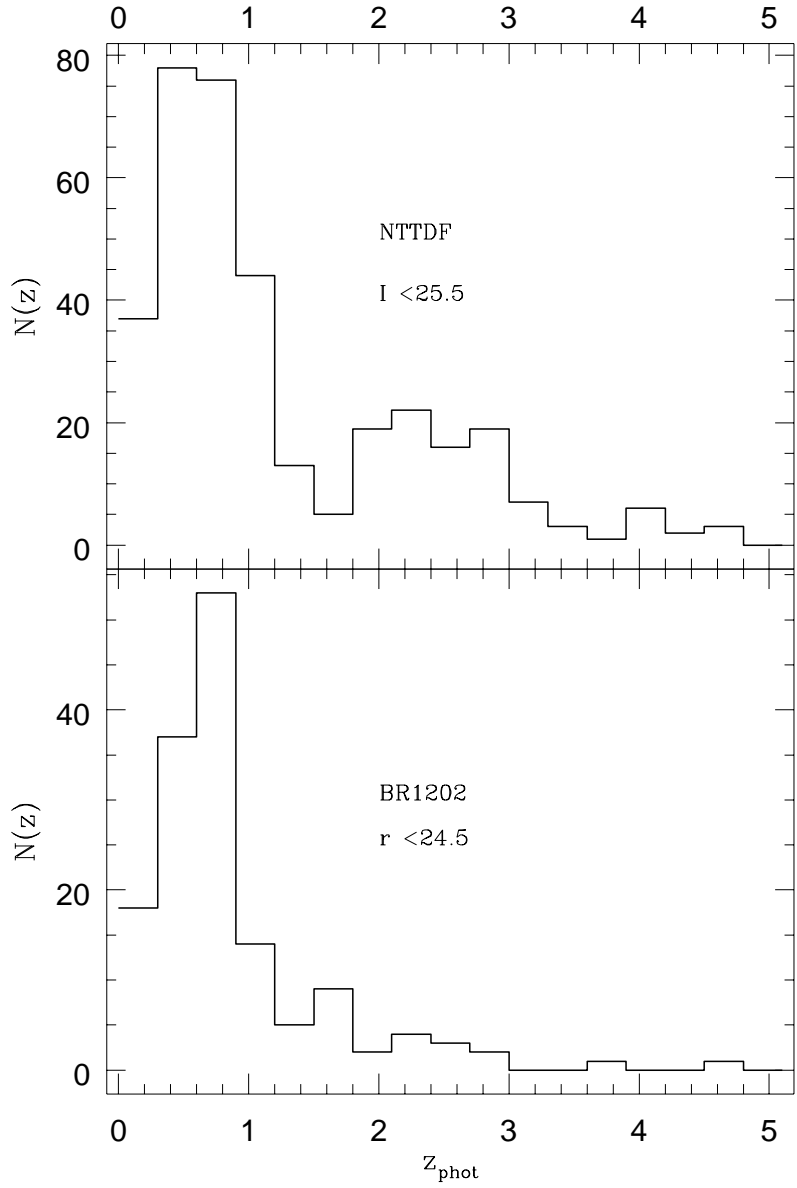}
\caption{
Redshift distribution in the NTT Deep Field (upper panel) and 
in the BR1202 field (lower). The samples are selected at different
magnitude limits (see legend).
}
\label{nz_br1202}
\end{figure}  

\clearpage
\begin{figure}
\plotone{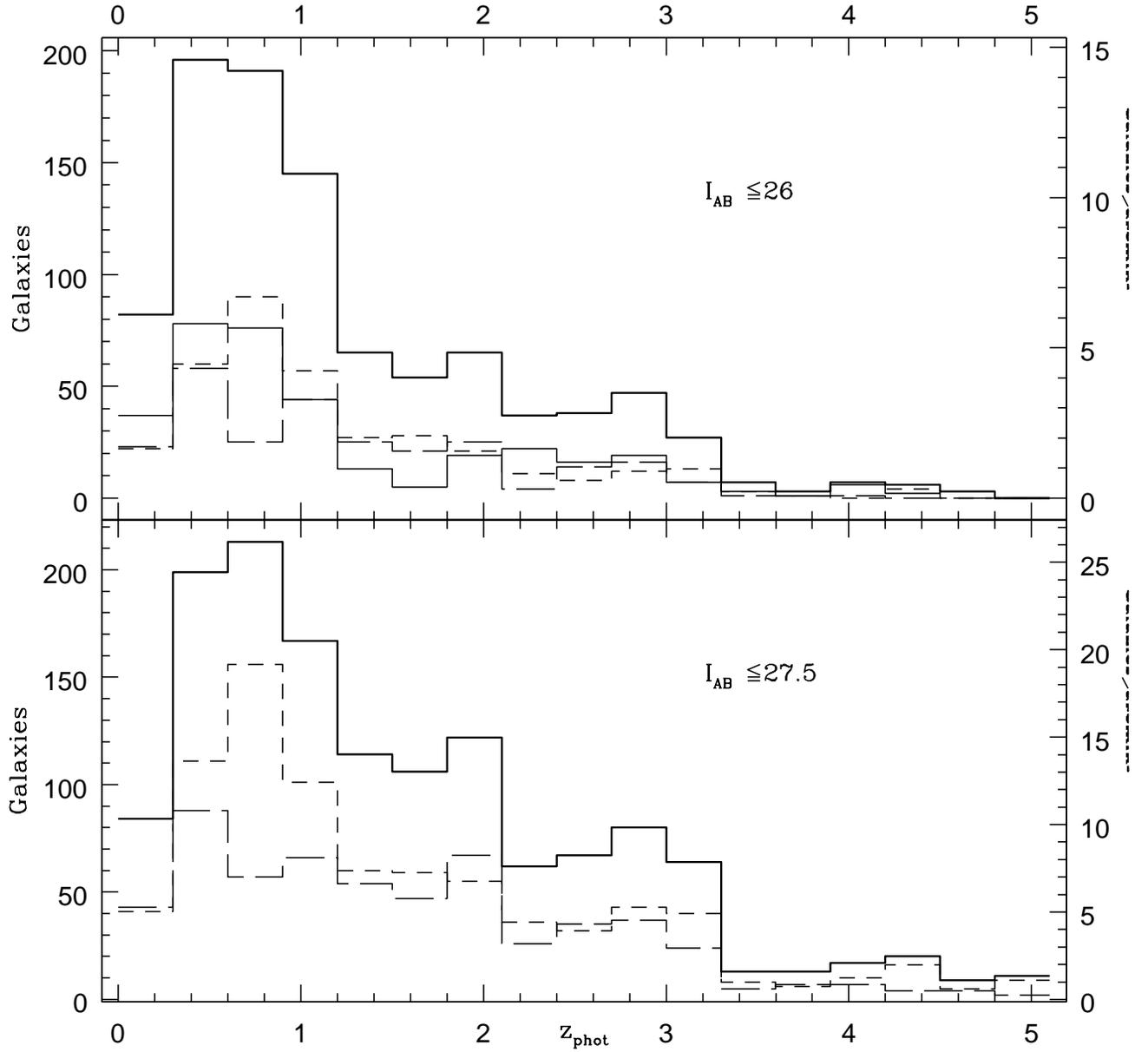}
\caption{
Redshift distribution in the faintest I--band selected catalogs.
Thick solid line is the total redshift distribution, thin solid line
is the NTTDF, dashed line is the HDF--N and long--dashed line is the
HDF--S. Left y--axis shows the number of galaxies in each bin,
while the right axis shows the surface density of galaxies in each bin in the
total sample (thick solid curve).
}
\label{nz_I}
\end{figure}  

\clearpage
\begin{figure}
\plotone{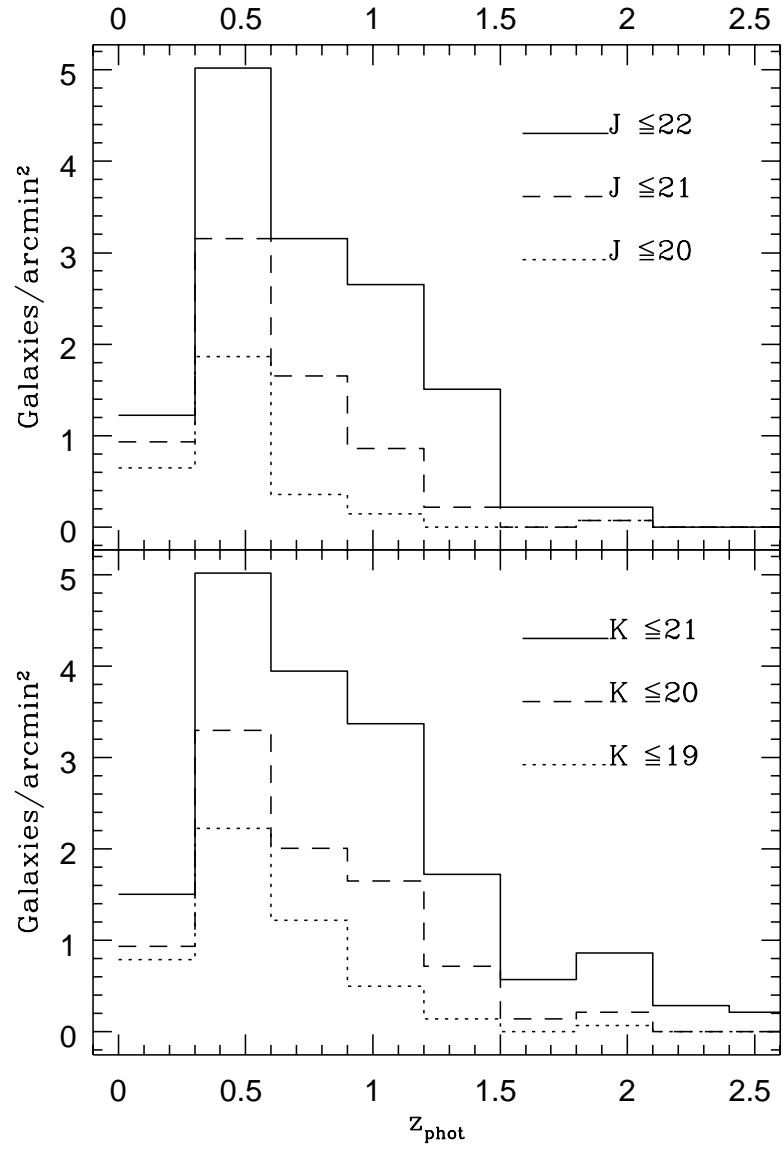}
\caption{
Redshift distribution in J and K$s$--band selected catalogs.
All curves are the sum of NNTDF, HDF--N and HDF--S.
Different magnitude limits are shown in the legends.
$J_{AB}$ and $K_{AB}$ magnitudes in the HDF catalogs have been converted to
Johnson applying $J = J_{AB} -0.87$ and $K = K_{AB} -1.84$ 
}
\label{nz_IR}
\end{figure}  

\clearpage
\begin{figure}
\plotone{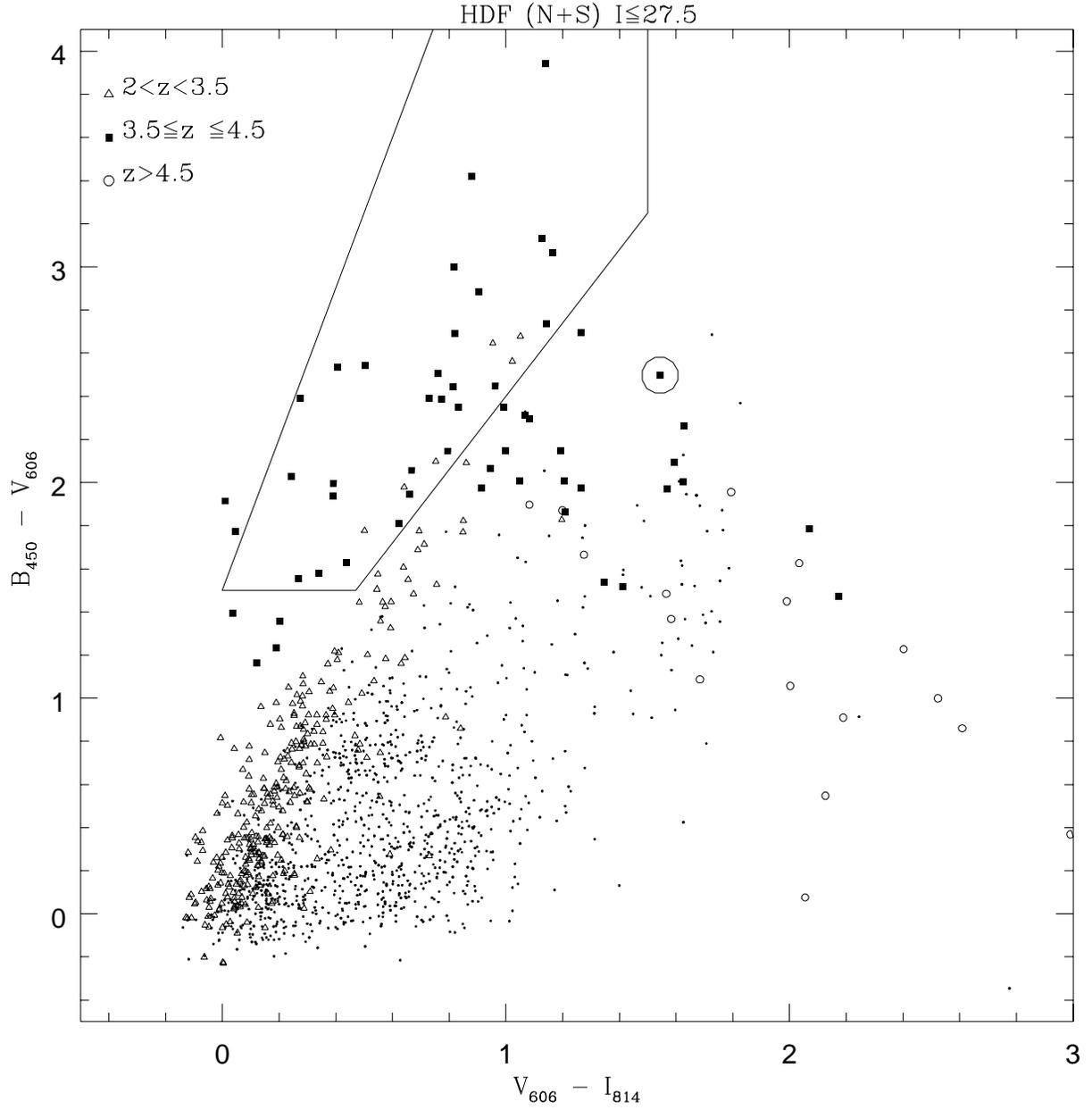}
\caption{ Position in the $B_{450}-V_{606}$ vs $V_{606}-I_{814}$ plane
of the galaxies in the HDF--N and HDF--S. Galaxies at photometric
redshifts $z_{phot}\geq2$ are shown with different symbols (see
legend). Flux upper limits are computed at 1$\sigma$ level.
The continous line defines the region proposed by M98
to select galaxies in the redshift range $3.5\leq z\leq 4.5$.  }
\label{madau}
\end{figure}  

\clearpage
\begin{figure}
\plotone{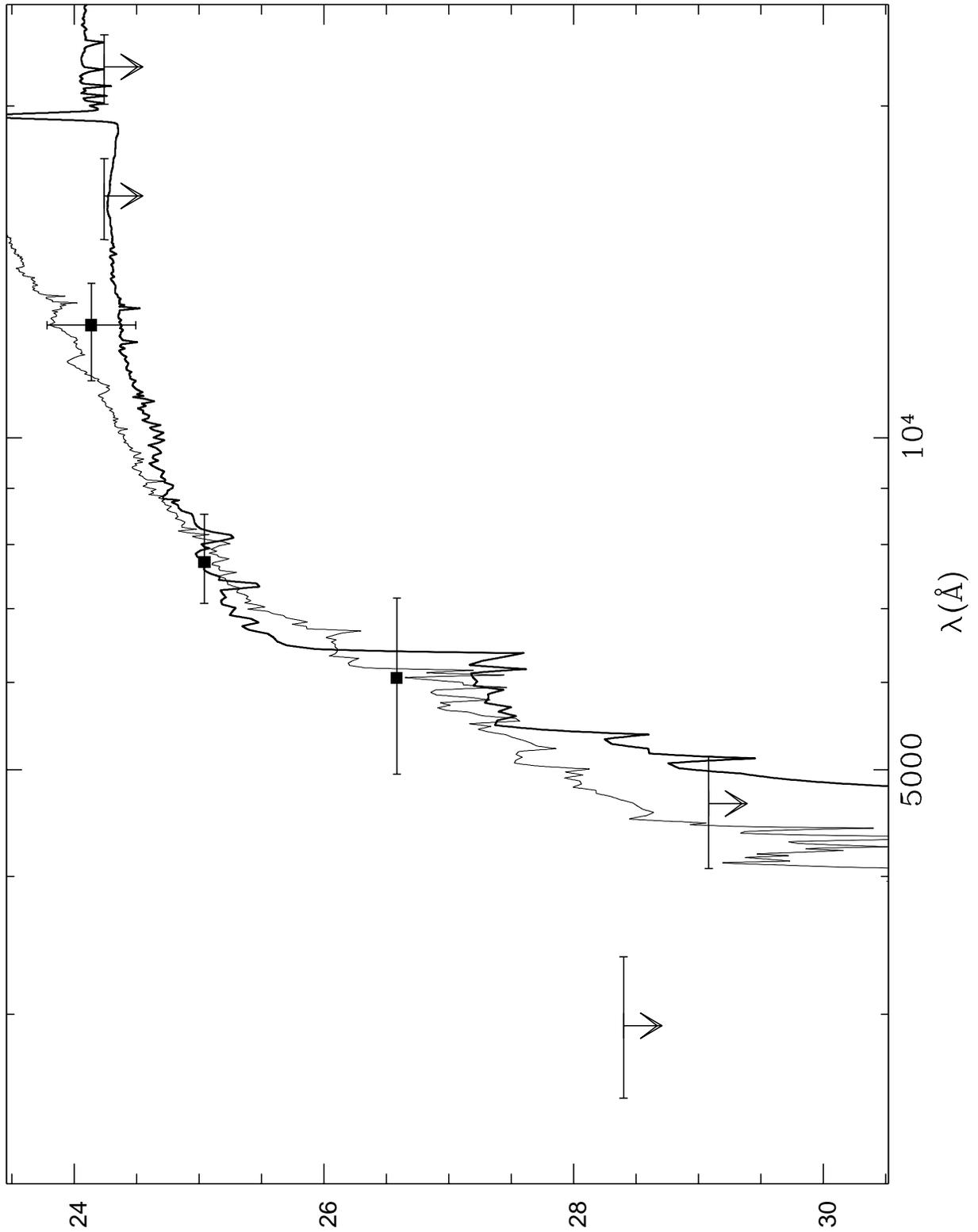}
\caption{Observed spectral distribution in the HDF--N filter set and
best--fitting spectra for HDF--N--3259--652, that has a spectroscopic
redshift of 4.58 and is assigned a photometric redshift $z =
4.26$. Solid line shows the best fitting spectrum at $z=4.26$ when the
IR bands are included, while the thin line shows the best fitting
spectrum at $z=0.57$ when only the 4 optical bands are used.}
\label{hdf173}
\end{figure}  

\clearpage
\begin{figure}
\plotone{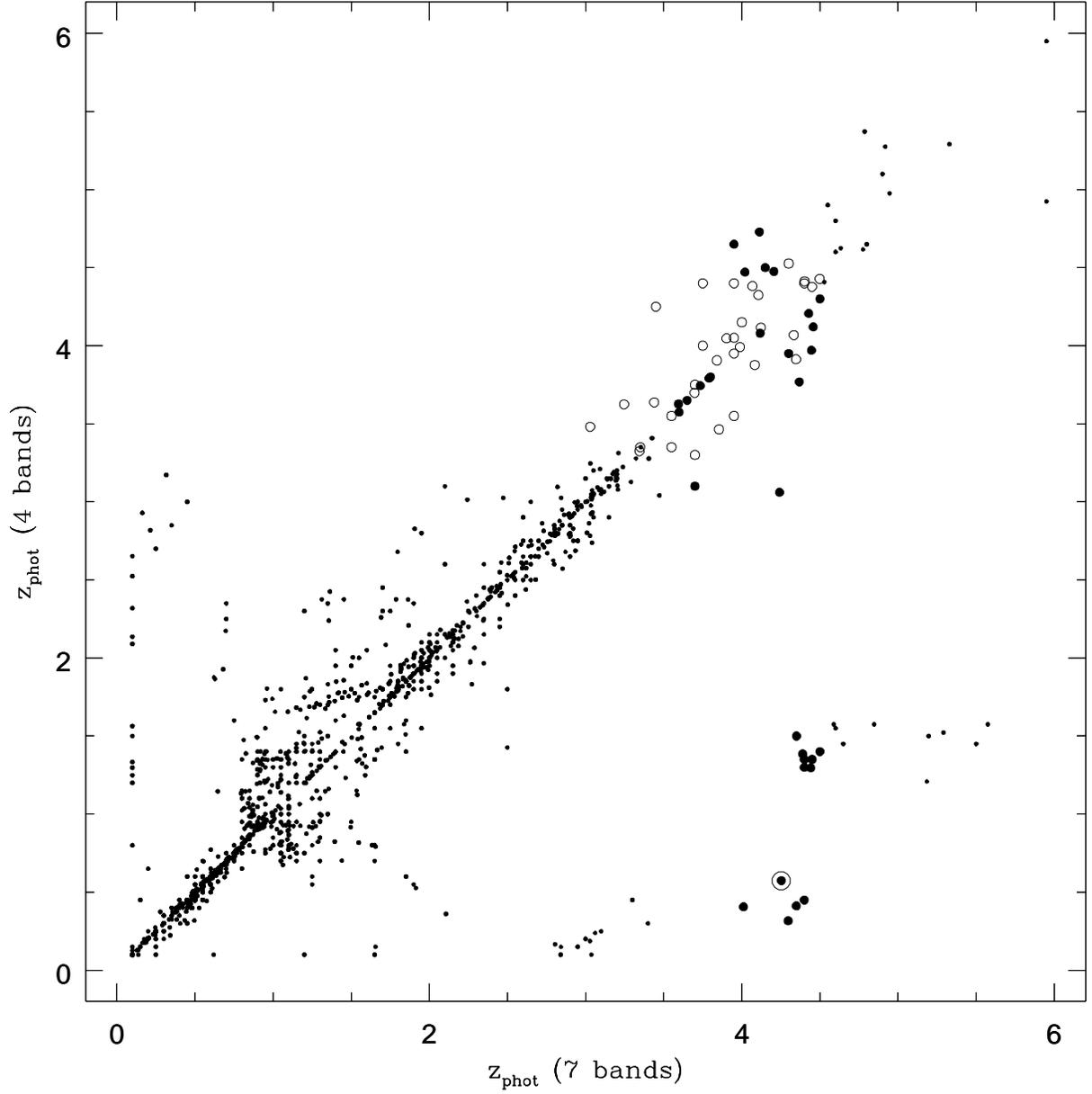}
\caption{Comparison between photometric redshifts in the HDF--N with 
all 7 bands (x--axis) and with  optical WFPC only (y--axis).
Large empty dots are galaxies lying within the M98 diagram at
$3.5<z<4.5$. Large filled  dot are galaxies outside the 
M98 selection criteria that are placed at $3.5<z_{phot}<4.5$
when all 7 bands are used. As in Fig.~\ref{hdf173}, 
HDF--N--3259--652 is marked by a large circle.
}
\label{noIR}
\end{figure}

\clearpage
\begin{figure}
\plotone{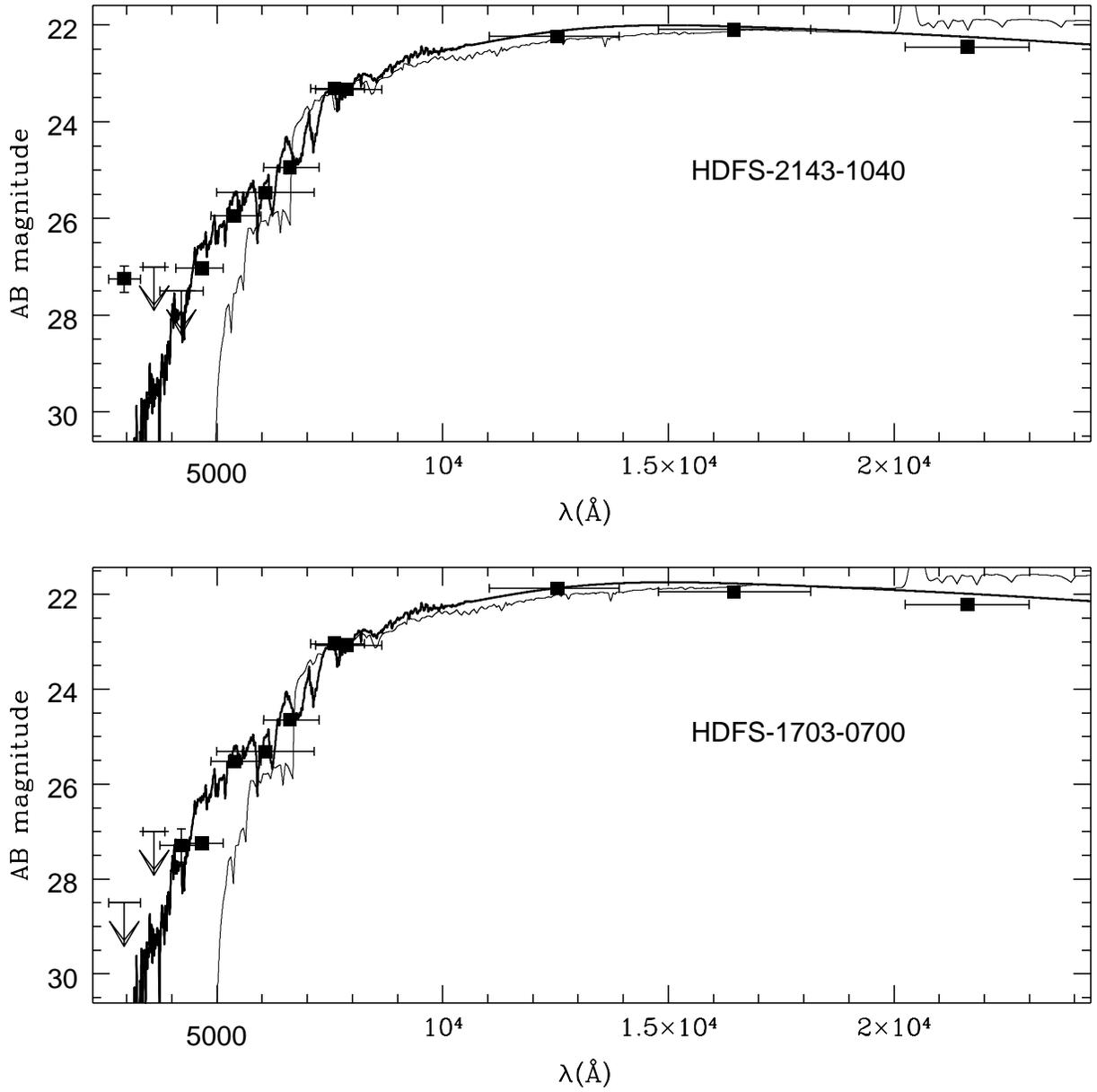}
\caption{Observed spectral distribution in the HDF--S filter set and
best--fitting spectra for HDFS--2143--1040 and HDFS--1703--0700.  Thin
solid line shows the best fitting spectrum at $z\simeq4.5$ when the IR
bands are included, while the thick line shows the best fitting
spectrum for an M star of the Pickles 1998 library.}
\label{mstar}
\end{figure}

\end{document}